\documentclass[twocolumn,showpacs,preprintnumbers,amsmath,amssymb]{revtex4}
\usepackage{dcolumn}% Align table columns on decimal point
\usepackage{bm}% bold math
\usepackage{latexsym}
\usepackage{amsfonts}
\usepackage[dvips]{graphicx}
\usepackage[usenames]{color}
\usepackage{makeidx}
\newcommand{\ket}{\rangle }
\newcommand{\bra}{\langle }

\newcommand{\ve}{\varepsilon}

\newcommand{\up}{\uparrow}
\newcommand{\dw}{\downarrow}

\newcommand{\Vect}[1]{\mbox{\boldmath$#1$}}

\begin{document}

%\preprint{APS/123-QED}
\title{Nonequilibrium superconducting and magnetic
phases in the\\ 
correlated electron system coupled to electrodes
}
\author{Takashi Oka$^{1,2}$ and Hideo Aoki$^1$}
\address{$^{1}$Department of Physics, University of Tokyo, Hongo, Tokyo 113-0033, Japan,\\
$^{2}$Theoretische Physik, ETH Z\"{u}rich, 8093 Z\"{u}rich, Switzerland}
\date{\today}
\begin{abstract}
\noindent 
A theory is presented for a nonequilibrium phase transition 
in the two-dimensional Hubbard model coupled to electrodes. 
Nonequilibrium magnetic and superconducting phase diagram is determined by the 
Keldysh method, where the electron correlation is treated in the 
fluctuation exchange approximation. 
The nonequilibrium distribution function
in the presence of electron correlation is evoked to 
capture a general feature in the phase diagram. 
\end{abstract}

\pacs{74.40.+k,05.30.-d,71.10.-w}
\maketitle

%%%%%%%%%%%%%%%%%%%%%%%%%%%%%%%%%%%%%%%%%%%
\section{ Introduction}
%%%%%%%%%%%%%%%%%%%%%%%%%%%%%%%%%%%%%%%%%%%
While our understanding of the physics of electron correlation 
has matured, there are still intriguing avenues that are yet to be 
fully explored.  One such avenue is 
strongly correlated electron 
systems in nonequilibrium situations. 
While there are a body of intense studies on 
nonequilibrium states in strong AC fields 
such as strong light sources 
that can  trigger photo-induced insulator-to-metal
transitions (see \cite{PIPTTokuraReview06} 
and refs therein), or nonequilibrium states in strong 
DC electric fields that 
can introduce pair-creation of electron and holes 
in dielectric 
breakdown \cite{tag,Oka2003}, 
here we pursue yet another
situation, where nonequilibrium 
states are conceived for an open, correlated 
electron system coupled to electrodes (Fig.\ref{fig:phasediagram} (a) inset).
Two effects are expected to arise from
the bias voltage $V$ across the electrodes.  
One is bi-carrier doping, i.e., 
electrons and holes are simultaneously doped, since two Fermi 
energies exist due to the two electrodes. 
Naively one might guess that this can make the system 
superconducting with Cooper pairs
formed by electrons or holes at half-filling, but this has to be tested.  
There is in fact the second effect, i.e., the electron-electron scattering 
in nonequilibrium that makes the originally sharp 
Fermi surface to be smeared. 
The smearing is expected to degrade 
magnetic orders \cite{MitraTakeiKimMillis06}, which
in our case implies that the smearing should 
act to reduce antiferromagnetic order. 
The natural question then is: will this also destroy the 
$d$-wave superconducting state? 

Here we study this problem, which is motivated by two recent experimental developments.  
One is the fabrication of functional 
structures with oxides \cite{OhtomoNature,Reyren07,Ueno08}.
In refs. \cite{OhtomoNature,Reyren07}, properties
such as superconducting transition in a
clean electron gas formed at an interface of two insulating 
oxides was studied, while  
Ueno {\it et al.} have succeeded in controlling the 
superconducting transition in an electrolyte-SrTiO$_3$ system
by changing the applied voltage. 
Nonlinear transport properties near the Mott transition at interfaces
have also been theoretically studied in 
\cite{OkamotoMillis04,OkaNagaosa,okamoto:116807}.

The second motivation comes 
from an experimental observation by Pothier 
{\it et al.} of a 
nonequilibrium electron distribution --- the double-step Fermi distribution --- 
in a mesoscopic copper wire attached to two electrodes \cite{Pothier97}. 
They showed that the step in the Fermi distribution is 
rounded due to electron scattering.  Such a 
smearing effect is expected to be even stronger 
in correlated electron systems,
so that it is theoretically 
imperative to develop a method 
for dealing with the nonequilibrium distribution of quasi-particles 
in a self-consistent manner in order to examine the nature 
of nonequilibrium phase transitions in 
correlated systems.  
Here we perform this by using the Keldysh method, 
while the interaction is treated within the 
fluctuation exchange approximation (FLEX) \cite{PhysRevLett.62.961,PhysRevB.43.8044}.
The superconductivity  transition is studied  
with the linearized Eliashberg equation.

We briefly comment on the past 
studies on superconductivity transition out of equilibrium.
In a pioneering work by Chang and Scalapino
 \cite{ChangScalapino78} who have solved the electron-phonon
model self-consistently, it was pointed out 
that nonequilibrium conditions such as irradiation of 
light can cause the quasiparticle distribution function to 
deform, and, under certain conditions, can lead to 
higher $T_c$ as observed in conventional $s$-wave superconductors 
\cite{Wyatt66,Kommers77}. 
In more recent attempts, critical 
properties near an insulator-superconductor 
transition were studied in 
\cite{Dalidovich04} followed by 
several authors \cite{MitraSC08,takei:165401}.

Here we adopt the Hubbard model, a 
prototype in the study of magnetism, superconductivity 
and other phase transitions in correlated electron systems.
In the two-dimensional square lattice near half-filling, 
the groundstate is the 
Mott insulator with an antiferromagnetic order
\cite{Imada1998}. 
When chemically doped with carriers (electrons
or holes), it is believed that 
Cooper pairs are formed with $d$-wave symmetry
and the system becomes superconducting,
\cite{PhysRevLett.62.961,PhysRevLett.72.1870,
PhysRevLett.72.1874,PhysRevLett.74.793} as 
also discussed phenomenologically in 
\cite{PhysRevB.42.167,PhysRevLett.67.3448,KurokiAoki96}.
So the question here is what happens in nonequilibrium.

%%%%%%%%%%%%%%%%%%%%%%%%%%%%%%%%%%%%%%%%%%%
\section{ Keldysh+FLEX method }
\begin{figure}[tbh]
\centering 
\includegraphics[width=8.5cm]{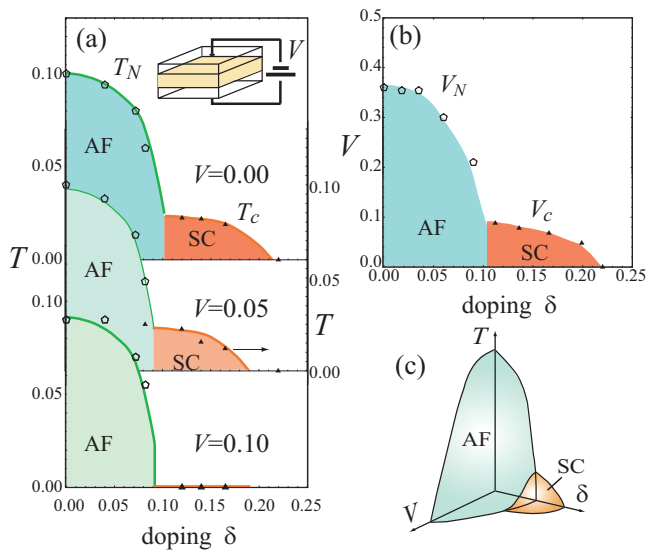}
\caption{(Color online) 
(a) The phase diagrams 
for various values of the bias voltage $V$ with 
AF (antiferromagnetic) and SC (superconducting) phases 
with $U/t=4.5$.  
Origins of the three panels are shifted for clarity, 
and shadings representing different phases are only a 
guide to the eye.  
Inset: Schematic sample (shaded) configuration with 
two electrodes.
(b) The zero-temperature phase diagram 
on the $(V, \delta)$ plane.   
(c) Schematic phase
diagram in the $(T, V, \delta)$ space. }
\label{fig:phasediagram}
\end{figure}

%%%%%%%%%%%%%%%%%%%%%%%%%%%%%%%%%%%%%%%%%%%
We consider a thin layer of strongly correlated material 
described by the two-dimensional Hubbard model which is 
coupled to electrodes.  
Here we have assumed  for simplicity the top and bottom electrodes 
(Fig.\ref{fig:phasediagram} (a) inset), since we want to 
single out the effect of different chemical potentials 
between the two electrodes, 
while a lateral attachment of 
the electrodes would cause a change in the 
spatial symmetry of the phases.  
The total Hamiltonian is then given by
\begin{eqnarray}
H&=&H_{\rm sys}+H_{\rm sys-electrode}+H_{\rm electrode},
\end{eqnarray}
where  
\begin{eqnarray}
H_{\rm sys}&=&-t\sum_{\bra i,j\ket,\sigma}(c^\dagger_{i\sigma}c_{j\sigma}+\mbox{h.c.})+U\sum_in_{i\up}n_{i\dw}
\end{eqnarray}
is the Hubbard Hamiltonian
with the hopping integral $t$ (taken to be the unit of energy 
hereafter) and the repulsive 
interaction $U$, while
\begin{eqnarray}
H_{\rm sys-electrode}&=&\sum_{i,\sigma,k,\gamma=1,2}
\left(V^k_\gamma c^\dagger_{i\sigma}a_{ik\sigma \gamma}+\mbox{h.c.}\right)
\end{eqnarray}
is the system-electrode coupling where we label the top (bottom) electrodes with $\gamma=1 (2)$, and $H_{\rm  electrode}$ the 
electrode Hamiltonian. 
The electrode electrons (created by $a^{\dagger}$) are free fermions
having correlators $\bra a_\gamma^\dagger a_\gamma\ket=f_\gamma$
with $f_\gamma$ the Fermi distribution function with electrode-dependent 
chemical potential $\mu_\gamma$.
The effect of the electrode can be taken into account
with the Schwinger-Dyson equation, 
where the self-energy, 
$
\Sigma^\alpha=\Sigma^\alpha_{\rm electrode}+\Sigma^\alpha_{\rm int}, 
$ consists 
of the contributions from the electrodes and those from the interaction.
Here $\alpha=r,a,<,>,K$ denote, respectively, 
the retarded, advanced, lesser,
greater, and Keldysh components (see, e.g., \cite{werner:035320,RammerBook}).
The electrode self-energy becomes 
\begin{eqnarray}
&&\Sigma^K_{\rm electrode}=2i\sum_{\gamma=1,2}\frac{\Gamma_\gamma}{2}
\tanh\frac{\omega-\mu_\gamma}{2T},\\
&&\Sigma^r_{\rm electrode}=-i\sum_{\gamma=1,2}\frac{\Gamma_\gamma}{2},
\end{eqnarray}
where $\Gamma_\gamma$ is the coupling strength between the 
system and the electrodes\cite{MitraTakeiKimMillis06,takei:165401}, 
$\mu_\gamma$  the respective chemical potential of 
the electrodes, 
and  the energy dependence in the density of states is neglected.
Here  the temperature $T$ of the two electrodes is kept to be 
the same, and we adopt $\Gamma_\gamma=0.001$.  
We note that 
if the coupling is too strong ($\Gamma_\gamma >\sim 0.1$), no
ordering takes place.

Nonequilbrium phase transitions can be studied 
by combining the Keldysh formalism with 
the FLEX to examine instabilities of the
nonequilbrium normal state against
magnetic and superconducting states.
The self-energy arising from the electron interaction 
is given, in nonequilbrium, by
\begin{eqnarray}
&&\Sigma^{>,<}_{\rm int}(\Vect{p},\omega)\\
&&=-i\int\frac{d\omega'}{2\pi}\int d\Vect{k}
P_{\rm eff}^{>,<}(\Vect{k},\omega')G^{>,<}(\Vect{p}-\Vect{k},\omega-\omega'),\nonumber
\end{eqnarray}
where $\Vect{p},\;\Vect{k}$ are momenta, $\omega$ the frequency,
and $N$ the number of $k$-points considered. 
The retarded component of the self-energy 
is determined from 
\begin{eqnarray}
\mbox{Im}\Sigma^r=\frac{1}{2i}\left(
\Sigma^>-\Sigma^<\right),
\end{eqnarray}
where the real part is obtained via Kramers-Kronig's relation.
Such relations between the 
lesser, greater and retarded components 
exist for other quantities as well.  
The fluctuation-mediated interaction, $P^{>,<}_{\rm eff}$, is given by
\begin{eqnarray}
P^{>,<}_{\rm eff}=U^2\mbox{Im}\left(
\frac{3}{2}\chi_s^{>,<}+\frac{1}{2}\chi_c^{>,<}
-\chi_{0}^{>,<}\right),
\end{eqnarray} 
where $\chi_{s}^\alpha (\chi_{c}^\alpha)$ represent 
the spin (charge) susceptibilities, 
whose retarded components are 
\begin{eqnarray}
\chi_s^r &=& \chi_{0}^r/(1-U\chi_{0}^r),\\
\chi_c^r &=& \chi_{0}^r/(1+U\chi_{0}^r).
\end{eqnarray}
Here $\chi_{0}$ is 
the irreducible susceptibility,
\begin{eqnarray}
&&\chi_0^{<,>}(\Vect{q},\omega)\\
&&=-i\int\frac{d\omega'}{2\pi}\int d\Vect{k}
G^{<,>}(\Vect{k},\omega')G^{>,<}(\Vect{k}+\Vect{q},\omega+\omega').\nonumber
\end{eqnarray}
The lesser and greater components of spin
and charge susceptibilties $\chi_{s,c}^\alpha$
are determined by
solving the Dyson equation.
For $\chi_s$, it is expressed by
\begin{eqnarray}
\chi_{s}^r&=&\chi_{s0}^r+U\chi_{s0}^r\chi_s^r,\\
\chi_s^{>,<}&=&\chi_{s0}^{>,<}+U\chi_{s0}^{>,<}\chi_s^a+
U\chi_{s0}^r\chi_s^{>,<}
\end{eqnarray}
obtained with aid of the Langreth rules
and can be solved by
\begin{eqnarray}
\chi_s^r&=&\chi_{s0}^r/(1-U\chi_{s0}^r),\\
\chi_s^{>,<}&=&\frac{\chi_{s0}^{>,<}}{(1-U\chi_{s0}^r)(1-U\chi_{s0}^a)}.
\end{eqnarray}
Similar expressions exist for $\chi_c$.
Finally,  Green's function is determined from the
self-energy through the Schwinger-Dyson equation, 
\begin{eqnarray}
(G^{r,a})^{-1}=(G_0^{r,a})^{-1}-\Sigma^{r,a},
\end{eqnarray}
for the retarded and advanced components, and
\begin{eqnarray}
G^{>,<}=G^r\Sigma^{>,<}G^a
\end{eqnarray} 
for the Keldysh component\cite{JauhoWingreenMeir94} 
with the bare Green's function
\begin{eqnarray}
G_0^{r,a}=(\omega-\ve_{\Vect{k}}\pm i\delta)^{-1}.
\end{eqnarray}
The process is repeated until a self-consistent 
solution is obtained. 
The {\it nonequilibrium distribution function} $f_{\rm eff}$
can be extracted\cite{tsuji} from the relation,
\begin{eqnarray}
G^K=(1-2f_{\rm eff})(G^r-G^a).
\label{eq:feff}
\end{eqnarray}
We seek for a self-consistent solution 
of the above equations with iteration until the self-energy 
converges.
In the calculation we take a 
64 $\times$ 64 grid 
for the square Brillouin zone,
while an almost logarithmic mesh \cite{PhysRevLett.74.793,PhysRevB.52.1297}
with 301 points for the $\omega$-axis is used.
We shall see that the distribution function $f_{\rm eff}$ deviates 
significantly from its 
non-interacting form (double step Fermi function),
\begin{eqnarray}
f_{\rm eff}^0 = [\Gamma_1f_{\rm FD}(\omega-\mu_1)+\Gamma_2f_{\rm FD}(\omega-\mu_2)]/(\Gamma_1+\Gamma_2),
\label{eq:f0}
\end{eqnarray}
(with $f_{\rm FD}$ being the Fermi-Dirac distribution) as an 
effect of the strong interaction.

The superconducting transition  
is studied in terms of the linearized 
Eliashberg equation, here extended to nonequilibrium.
To this end, we iteratively ($i=1,2,\ldots$) obtain
the anomalous self-energy $(\phi^\alpha_i)$ and anomalous 
Green's function $(F_i^\alpha)$ 
using $\Sigma^\alpha$, $\chi_{s,c}^{\alpha}$ obtained 
in the previous step.  
With a random initial guess for $\phi_1^r$, 
Green's function is determined from 
the linearized Nambu-Gor'kov equation, 
\begin{eqnarray}
F^r_i &=& \frac{\phi^r_i}{(\omega Z)^2-(\ve_{\Vect{k}}+X)^2},\\
\omega Z&=&\omega-[\Sigma^r(\omega)-(\Sigma^{r}(-\omega))^*]/2,\\
X&=&[\Sigma^r(\omega)+(\Sigma^{r}(-\omega))^*]/2.
\end{eqnarray}  
Then the Keldysh component is calculated with the 
generalized distribution function, 
\begin{eqnarray}
F^{K}_i=(1-2f_{\rm eff})(F^r_i-F^a_i).
\end{eqnarray}
We assume here that the distribution for the
anomalous component is the same as that for 
the normal component.
Finally, we plug this into the Eliashberg equation, 
\begin{eqnarray}
&&\phi^{>,<}_{i+1}(\Vect{p},\omega)\\
&&=-i\int\frac{d\omega'}{2\pi}\int d\Vect{k}P_{\rm sing}^{>,<}(\Vect{k},\omega')F^{>,<}_i(\Vect{p}-\Vect{k},\omega-\omega'),\nonumber
\end{eqnarray}
where the effective interaction in the spin-singlet channel is 
\begin{eqnarray}
P^{>,<}_{\rm sing}=U^2\mbox{Im}\left(
\frac{3}{2}\chi_s^{>,<}-\frac{1}{2}\chi_c^{>,<}\right).
\end{eqnarray}
The eigenvalue of the linearized Eliashberg equation is obtained as 
$\lambda=\lim_{i\to \infty}||\phi_{i+1}^r||/||\phi_{i}^r||$, 
where $||\phi_i^r||=(\int d\omega \int d\Vect{p}\phi^r(\Vect{p},\omega)|^2)^{1/2}$ is the norm. The superconducting transition takes place
when $\lambda$ exceeds unity.

Before moving on to the results, we comment on the 
applicability of the FLEX on the magnetic transition. 
In our formalism, we have used the RPA expression 
for the susceptibility combined with the 
FLEX following Ref.\cite{PhysRevLett.62.961}.
In equilibrium, this approximation 
gives a phase diagram for magnetic and 
superconducting transitions where the 
superconducting phase cuts the AF dome. 
The formalism has limitations in that  
(a) it cannot describe the Mott physics or the pseudo-gap,  and 
(b) the magnetic transition is not recovered when 
one uses the FLEX spin susceptibility instead of the 
RPA form (see \cite{PhysRevB.43.8044}). 
Thus the approach developed here should be considered to 
be limited to the weak-coupling regime.

%%%%%%%%%%%%%%%%%%%%%%%%%%%%%%%%%%%%%%%%%%%%%
\begin{figure}[tbh]
\centering 
\includegraphics[width=8.5cm]{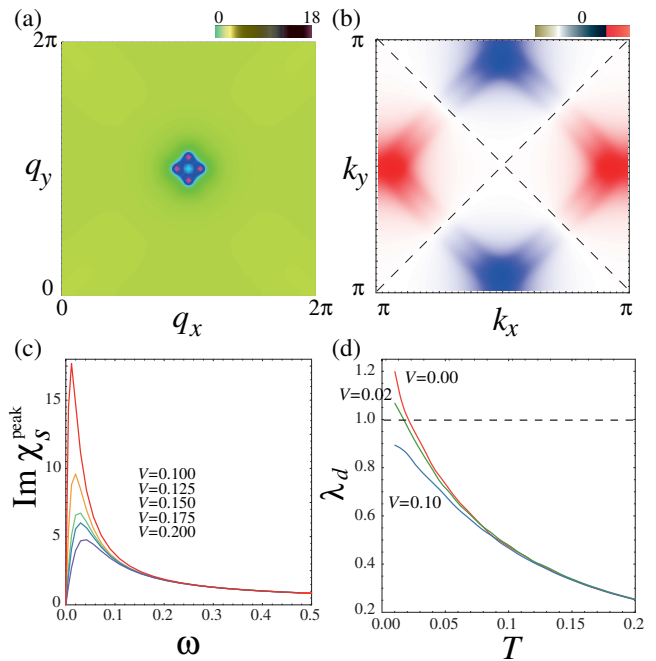}
\caption{(Color online) 
Spin susceptibility 
$\mbox{Im}\,\chi_s(\Vect{q},\omega)$ (a)  and superconducting gap function
$\mbox{Re}\,\phi(\Vect{k},\omega= 0)$ (b) are color-coded versus momentum 
for a bias $V=0.1$ above the critical value, 
with the doping level $\delta=0.14$, 
$U=4.5$, and $\mu=-0.35$.  
Dashed lines in (b) represent the nodes.  
(c) The peak value 
$\mbox{Im}\,\chi_s^{\rm peak}(\Vect{q},\omega)$  versus $\omega$ 
for $V=0.1 - 0.2$ from top to bottom 
for $\Vect{q}=(\pi,1.1\pi)$.   $T=0.002$ for (a)-(c).
(d) The temperature dependence of the Eliashberg 
eigenvalue $\lambda_d$ for 
the $d$-wave pairing for $V=0.0 - 0.1$ 
from top to bottom.
}
\label{fig:statedata}
\end{figure}
%%%%%%%%%%%%%%%%%%%%%%%%%%%%%%%%%%%%%%%%%%%%

%%%%%%%%%%%%%%%%%%%%%%%%%%%%%%%%%%%%%%%%%%%%%
%%%%%%%%%%%%%%%%%%%%%%%%%%%%%%%%%%%%%%%%%%%%

%%%%%%%%%%%%%%%%%%%%%%%%%%%%%%%%%%%%%%%%%%%
\section{ Nonequilibrium phase transition }
%%%%%%%%%%%%%%%%%%%%%%%%%%%%%%%%%%%%%%%%%%%
We have applied the above formalism 
to obtain the nonequilibrium phase diagram for the 
two-dimensional (square lattice) Hubbard model 
attached to two electrodes by numerically solving the
equations self-consistently.  
In equilibrium the phase diagram within FLEX as 
obtained in \cite{PhysRevLett.62.961} has an 
antiferromagnetic phase when the doping level $\delta=1-n$
is small, which is taken over by a $d$-wave superconductivity as 
$\delta$ is increased.   So the interest is how the nonequilibrium 
situation modifies these.   We first plot 
in Fig. \ref{fig:statedata}(a) 
the spin susceptibility 
$\mbox{Im}\,\chi_s(\Vect{q},\omega)$ 
for $V=0.1$ and a doping level $\delta=0.14$ 
The result shows that the antiferromagnetic fluctuation
remains strong near half-filling, 
for which we have four  incommensurate peaks around 
$\Vect{q}=(\pi,\pi)$ in $k$-space, as in equilibrium.   
The effect of increased bias is that the peak height is reduced, 
and the peak position on energy axis shifts upwards as displayed 
in Fig. \ref{fig:statedata} (c), 
where $\mbox{Im}\,\chi_s^{\rm peak}(\Vect{q},\omega)$ 
for $\Vect{q}=(\pi,1.1 \pi)$ is plotted.  
We notice that no features such as dip or
hump appear around $\omega\sim V$.
The dominant superconducting solution in Fig. \ref{fig:statedata}(b) is again similar to the 
equilibrium case, that is, the $d$-wave gap has
the largest  $\lambda_d$ for the 
linearized Eliashberg equation.   
However, the critical temperature $T_c$ at which $\lambda_d$ 
reaches unity 
depends on $V$, as shown by the 
temperature dependence of $\lambda_d$  plotted 
in  Fig. \ref{fig:statedata}(d).   
So the bias $V$ reduces $T_c$, until finally 
the superconducting 
state no longer exists even at zero temperature 
when the bias becomes too strong.  
We define this as the {\it critical bias} $V_c$. 
For the region of the band filling for which 
the antiferromagnetic order  
dominates over the superconducting state, 
we can define the bias-dependent N\'{e}el temperature 
$T_N$ as the temperature at which 
the spin susceptibility diverges\cite{NeelDef}.
The spin susceptibility is 
reduced as the bias in increased, until the 
antiferromagnetic order vanishes even 
at zero temperature beyond  the ``critical N\'{e}el bias" $V_N$.  
The doping dependence of the N\'{e}el bias and the critical 
temperatures for a fixed bias is shown in Fig. \ref{fig:phasediagram} (a). 
We can see that, while the antiferromagnetic 
(AF)  phase is relatively persistent,
the superconducting (SC) region rapidly shrinks 
with the bias $V$ and disappears at $V\simeq 0.1$.

The phase diagram at zero-temperature is plotted on the $(V, \delta)$ 
plane in Fig. \ref{fig:phasediagram} (b).  
The N\'{e}el bias, peaked at the undoped point with $V_N\simeq 0.36$,
decreases with the doping, 
and the AF phase is replaced with the SC 
phase around $\delta\simeq 0.1$ with a maximum
critical bias for SC $V_c\simeq 0.1$.  
As we further increase the doping, the SC 
phase finally disappears. 
 Figure  \ref{fig:phasediagram} (c) schematically summarizes 
the phase transitions in the $(T, V, \delta)$ space.

%%%%%%%%%%%%%%%%%%%%%%%%%%%%%%%%%%%%%%%%%%%
\section{ Nonequilibrium distribution function}
%%%%%%%%%%%%%%%%%%%%%%%%%%%%%%%%%%%%%%%%%%%
As was experimentally found in a tunneling
measurement in a mesoscopic wire of copper 
by Pothier {\it et al.},\cite{Pothier97} 
the nonequilibrium electron distribution 
becomes smeared from the simple, double-step 
Fermi distribution $f_{\rm eff}^0$ 
due to electron scattering. 
In correlated materials with a strong 
electron-electron interaction, 
we expect a greater smearing effect to take place. 
Indeed, as we shall reveal below, 
the key feature to understand the nonequilibrium 
phase diagram for the open Hubbard model may be 
captured by the way in which the nonequilibrium distribution function
is rounded by the interaction effect. 

%%%%%%%%%%%%%%%%%%%%%%%%%%%%%%%%%%%%%%%%%%%%%
\begin{figure}[tbh]
\centering 
\includegraphics[width=8.cm]{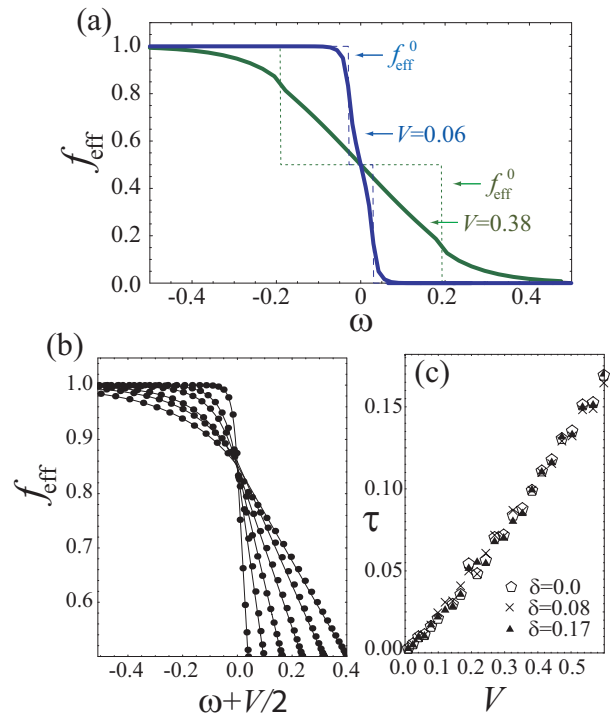}
\caption{(Color online) 
(a) Nonequilibrium distribution function
for two values of the bias, $V=0.06$ 
or $V=0.38$, at half filling ($\delta=0$).
Dashed lines 
are the noninteracting distribution function $f_{\rm eff}^0$. 
(b) Nonequilibrium distribution function (dots) against $\omega$ 
in the $\omega<-V/2$ region 
for $V=0.08,\;0.19,\;0.32,\;0.47,\;0.63,\;0.80$ from the top, 
where curves represent 
a fit with eq.(\ref{eq:fit}). 
(c) The smearing parameter $\tau$ against the bias $V$
for various values of $\delta$ and 
fixed $U=4.5$ and $T=0$.  
Fitting errors are smaller than the size of each symbol.
}
\label{fig:distribution}
\end{figure}
%%%%%%%%%%%%%%%%%%%%%%%%%%%%%%%%%%%%%%%%%%%%

Figure \ref{fig:distribution} (a) plots the effective 
distribution $f_{\rm eff}$ defined in eq.(\ref{eq:feff}) 
obtained self-consistently for two values of the bias $V$.  
The temperature in the electrodes, 
hence in $f_{\rm eff}^0$, is set to zero. 
If we compare the result with the corresponding noninteracting distribution
function $f_{\rm eff}^0$ (eq.(\ref{eq:f0})) 
(dashed lines), $f_{\rm eff}$ is seen to 
significantly deviate from $f_{\rm eff}^0$.   
More importantly, we find here 
that the {\it effective temperature approximation 
breaks down}, that is, we cannot fit $f_{\rm eff}$
to $f_{\rm eff}^0$ 
with the temperature as a fitting parameter. 
Instead, the best fit to the data is given by 
\begin{eqnarray}
f_{\rm eff}^{\rm fit}=\left\{
	\begin{array}{cl}
1-\alpha e^{(\omega+V/2)/\tau},&\omega<-V/2\\
-(1-2\alpha)\omega/V+1/2,&-V/2\le\omega<V/2\\
\alpha e^{-(\omega-V/2)/\tau},& V/2\le\omega
\end{array}
\right.
\label{eq:fit}
\end{eqnarray}
where $\alpha$ and $\tau $ are the 
fitting parameters. 
The parameter $\tau$
having the dimension of energy 
represents the extent to which the distribution is 
smeared from the double-step function. 
%We call this the smearing parameter so as to stress the fact that the effective temperature approximation does not work.
We have found in Fig. \ref{fig:distribution} (b) 
that the fitting function 
eq.(\ref{eq:fit}) is adequate 
in the present open Hubbard model
in that all the data for various values of the 
parameters ($V, \Gamma, U, \delta, \ldots$) 
are reproduced within the numerical errors.  
If we specifically plot the bias-dependence of the 
smearing parameter in Fig. \ref{fig:distribution} (c), we can 
see that they fall upon an  universal curve. 
When $V$ is small, one can approximate this with a
linear relation, 
\begin{eqnarray}
\tau\propto V.
\label{tauproptoV}
\end{eqnarray}
The proportionality constant depends on 
the interaction strength 
$U$ and the coupling 
$\Gamma$ to the electrodes, but not on the filling $\delta$ 
as seen from the figure. 
The constant is reduced when 
the coupling to the electrode becomes stronger.

From the viewpoint of the smeared distribution, 
we can conceive the bias-driven
phase transitions in the following way.  
We have seen in Fig. \ref{fig:phasediagram} (b) 
that the AF (SC) orders 
die out at $V\simeq 0.4$ ($V\simeq 0.1$) respectively.  
In terms of eq.(\ref{tauproptoV}), 
these values correspond to the smearing 
parameters $\tau\simeq 0.1$ ($\tau\simeq 0.02$).  We can then 
note that these values are similar to 
the highest N\'{e}el (critical) temperatures 
in the zero bias phase diagram 
(Fig. \ref{fig:phasediagram} (a), upper panel).  
Thus, the transition takes place 
when  the smearing parameter $\tau$ attains 
a value ({\it depth} of each phase 
in the phase diagram 
in Fig. \ref{fig:phasediagram} (c) as translated to $\tau$) 
that is similar to the transition 
temperature ({\it height} in the same phase diagram).  
AF spin fluctuations are suppressed 
in finite bias voltages in this manner, 
which is similar to what happens in itinerant electron magnets
\cite{MitraTakeiKimMillis06}. 

%%%%%%%%%%%%%%%%%%%%%%%%%%%%%%%%%%%%%%%%%%%
\section{ Discussion}
%%%%%%%%%%%%%%%%%%%%%%%%%%%%%%%%%%%%%%%%%%%
We have obtained a nonequilibrium phase diagram for the 
two-dimensional Hubbard model, and pointed out the possibility 
of controlling the phases in strongly correlated
heterostructures (i.e., electrode-system-electrode) by external bias.
 Both of AF and SC regions shrink with the 
bias $V$, which we attribute to the smearing of the nonequilibrium 
distribution function.  
While the smearing can be reduced 
if we make the system more strongly coupled to the 
electrodes (in e.g. a  thinner sample), 
this will lead to the destruction of orders because 
a larger coupling $\Gamma$ to electrodes 
will make the spin fluctuations weaker. 
Thus we conclude the smearing of the distribution 
function is an important property characterizing 
correlated electron systems
out of equilibrium, and an experimental 
verification of this should be interesting. 
We have to make a caution that 
FLEX employed here has limitations in that it 
ignores the vertex correction, and cannot address, due to its 
weak-coupling nature, the
behavior close to the Mott insulator point, as mentioned. 
Effects of electrodes (on e.g. the pairing symmetry) when they are attached laterally are also intriguing.   
A more ambitious future problem is a possibility of 
bi-carrier induced superconductivity in nonequilibrium, for which the 
present formalism may serve as a starting point.

TO wishes to thank Thomas Dahm and Yoichi Yanase for helpful advices.  
%HA acknowledges JST-TRIP 

%\bibliographystyle{apsrev.bst}
%\bibliographystyle{apsrmp.bst}
%\bibliographystyle{unsrt.bst}
%\bibliography{c:/Physics/ref}

\begin{thebibliography}{27}
\bibitem{NeelDef}
In numerical calculations 
the divergence is rounded, 
and we adopt an often employed 
criterion for 
the transition, Max[$\mbox{Im}\,\chi_s(\Vect{q},\omega)]>50$.
\end{thebibliography}
%\printindex

\end{document}